\documentclass[aps,nofootinbib]{revtex4}
\usepackage{amsmath}
\usepackage{graphicx}

\def\rar{\mathop{\longrightarrow}\limits}

\begin{document}
\draft
\title{On the effectiveness of effective field theory in peripheral 
nucleon-nucleon scattering}
\author{Michael C. Birse and Judith A. McGovern}
\address{Theoretical Physics Group, Department of Physics and Astronomy\\
University of Manchester, Manchester, M13 9PL, UK\\}

\begin{abstract}
Peripheral nucleon-nucleon scattering is analysed in the framework of an 
effective field theory. Distorted-wave methods are used to remove the effects
of one-pion exchange. Two-pion exchange and recoil corrections to one-pion 
exchange are then subtracted pertubatively. This removes all contributions up 
to order $Q^3$ in the chiral expansion. We have applied this to the $^1D_2$,
$^1F_3$ and $^1G_4$ waves, using phase shifts from various
partial-wave analyses by the Nijmegen group. In regions where these
analyses agree we find no evidence for a breakdown of the chiral expansion.
One of the terms in the effective short-range potential, the leading one
in the $^1D_2$ channel, is larger than might be expected, but in general these
terms have momentum scales of about 300--400 MeV. We also see hints of isospin 
breaking in the $\pi N$ couplings.
\end{abstract}
\maketitle

\vskip 10pt

\section{Introduction}

Following the suggestion of Weinberg \cite{wein} and the pioneering work of
Ordo\~nez, Ray and van Kolck \cite{orvk}, there has been much interest in 
using chiral perturbation theory (ChPT) to calculate long-range contributions 
to the nucleon-nucleon ($NN$) force.\footnote{For a recent review of effective 
theories of $NN$ scattering, see Ref.~\cite{BvK02}.}
These have concentrated on two-pion exchange (TPE)
\cite{kbw,kgw,egm1,nij99,friar,brr,rich,kais,em1,nij03,egm2}, 
although work has also started on three-pion exchange \cite{kais3,egm3}. In an
effective field theory like this, the residual short-distance interactions 
are represented by energy- or momentum-dependent contact interactions.
These can be expanded in powers of the ratios between momenta or the pion 
mass and the scale of the physics which is responsible for these short-range
interactions, for example the mass of the $\rho$ meson. Provided these ratios 
are small enough, this expansion should converge rapidly.

Nucleon-nucleon scattering in peripheral waves provides the best place to 
look for clear signals of the long-range TPE force. Various groups have 
done so \cite{kbw,kgw,egm1,rich,egm2} but they often find quite 
large deviations from the ``experimental" phase shifts deduced from 
partial-wave analyses \cite{nijnn,said}. This has led some to resort to 
introducing phenomenological cut-offs \cite{egm1,egm2} or additional degrees 
of freedom \cite{kgw}. 

Recently the Nijmegen group has started carrying out PWA's which include chiral 
TPE \cite{nij99,nij03} and they find that this makes significant improvements 
to their fits. The large differences between their results and available ChPT 
predictions have led to a claim that the fits cannot yield reliable values for 
parameters in the TPE potential \cite{em2}. However it is hard to make a 
direct comparison between the Nijmegen analyses and ChPT because they are done
with a coordinate-space cut-off of 1.4 fm or larger. This means that 
short-distance parameters in the fits have to play two roles. As well as 
parametrising true short-range physics they also have to correct for 
artefacts introduced by the cut-off. Hence one cannot immediately tell from
the Nijmegen parametrisation whether the residual short-range interactions are 
consistent with an effective field theory.

Distorted wave (DW) methods can be used to extract the effects of a known 
long-range interaction from two-body scattering, leaving a residual
scattering amplitude which can be analysed using the techniques of
effective field theory \cite{bb}. For systems with bound states close to threshold, 
such as $NN$ $S$-waves, this is equivalent to a DW version of the effective-range
expansion. In peripheral waves, where the scattering is weak, the effective
short-range potential is essentially just an expansion of the residual 
$K$-matrix in powers of the energy. 

Related approaches can be found in Refs.~\cite{brr,pr} where a variable-phase 
method is used to construct the DW's and a radial cut-off is imposed at
some small radius. Ballot {\it et al.}~\cite{brr} include TPE but do 
not attempt to extract a residual interaction strength, only examining the 
sensitivity of their results to the cut-off radius. (See also Ref.~\cite{rich} 
for a similar treatment.) Pavon Valderrama and Ruiz Arriola \cite{pr} keep only 
OPE and parametrise the residual interaction via a boundary condition at their 
cut-off radius. This is constructed to reproduce exactly the effective-range 
expansion up to fourth order in the momentum.

Peripheral partial waves are of particular interest for testing $NN$ potentials
from ChPT, since the nucleons do not feel the very strong attraction 
that causes complications in the $S$-waves \cite{BvK02}. However phase shifts in 
a wave with orbital angular momentum $L$ typically grow like the $L$-th power of 
the energy. This means that quite small differences between the short-distance
behaviour of potentials can lead to phase shifts that look very different
at energies $\sim 2m_\pi$. Conversely, at low energies, small differences
between phase shifts can be hard to disentangle. In particular, differences
between the available PWA's in this region can be comparable in size to the 
TPE effects of interest, but this is not always obvious in simple plots of
phase shifts. By applying the DW method to empirical phase shifts, we are able 
to obtain a residual interaction strength that is much less energy dependent than 
the phase shifts themselves. This has the advantage of providing a quantitative
measure of the strength of the missing physics and of its energy dependence. 
It also makes much clearer the regions where the PWA results are not reliable.

Here we use the DW method to examine $NN$ scattering in the $^1D_2$, $^1F_3$ 
and $^1G_4$ partial waves using the phase shifts from five available Nijmegen 
analyses: PWA93, and the Nijmegen I, Nijmegen II, Reid93 and ESC96 potentials 
\cite{nijnn}. All of these include the long-range OPE potential but otherwise
they parametrise the data quite differently. For example, the PWA imposes an 
energy-dependent boundary condition at a cut-off radius of 1.4 fm, while the 
potentials are energy-independent but may be either local or nonlocal.
All of them have been fitted with similarly good $\chi^2$ to to the $NN$ data 
available in 1993, and so they can be regarded as alternative PWA's. 
Any differences between these analyses should be taken as 
indications of the size of the systematic errors associated with them. The 
reason for concentrating initially on the spin-singlet waves is that the 
OPE potential is less singular than the centrifugal barrier and hence no 
regularisation is needed to construct the distorted waves. The need for some 
additional regulator in the triplet waves means that the short-range potential 
would have to play two roles, correcting for artefacts of the regulator as 
well as parametrising short-range physics, and so it would be harder to 
interpret.

The terms in the interaction can be classified according to a chiral expansion 
in powers of $Q$, where $Q$ denotes a factor of either a momentum or a pion 
mass \cite{BvK02}. In this counting the leading OPE potential is of order
$Q^0$. We use the TPE potential given in Refs.~\cite{kbw,nij99}. This
includes terms up to order $Q^3$ \cite{kbw}. 
As discussed by Friar \cite{friar}, this potential should be used in 
conjunction with an OPE potential that has the usual nonrelativistic
form (see Eq.~(\ref{eq:npope}) below) multiplied by $M/E$ where $E$ 
is the on-shell energy of one nucleon. This factor generates recoil 
corrections to OPE which start at order $Q^2$. Other corrections to OPE which 
might arise from higher-order $\pi N$ vertices can all be absorbed in the 
on-shell $\pi N$ couplings or $NN$ contact terms (as discussed in 
Ref.~\cite{egm1}) or in higher-order terms in the expansion of $M/E$.

The order-$Q^2$ recoil correction to OPE plus TPE terms from 
Ref.~\cite{kbw,nij99} therefore provide the complete long-range
potential at orders $Q^2$ and $Q^3$, with one exception:
we use a single $\pi N$ coupling constant. At this order there
could be isospin breaking in the $\pi N$ couplings 
\cite{vkfg}, a point to which we shall return later.
By using the DW method to extract all iterations of OPE and 
then subtracting these OPE and TPE terms, we are able to remove
all contributions to the scattering up to order $Q^3$ in the chiral
expansion.

\section{Distorted-wave method}

The starting point for the approach of Ref.~\cite{bb} is the $K$-matrix which 
describes scattering between distorted waves of the known long-range potential,
in our case OPE. On shell, this matrix $\tilde K(p)$ is related to the 
observed phase shift $\delta(p)$ by 
\begin{equation}
\tilde K(p)=-{4\pi\over Mp}\tan\Bigl(\delta(p)
-\delta_{\scriptscriptstyle\rm OPE}(p)\Bigr),
\end{equation}
where $p$ is the on-shell relative momentum in the c.m.\ frame. Here  
$\delta_{\scriptscriptstyle OPE}(p)$ denotes the phase shift for the 
lowest-order OPE potential. In the spin-singlet $np$ channels this potential 
has the form
\begin{equation}\label{eq:npope}
V^{(0)}_{\scriptscriptstyle\rm OPE}(r)=-f_{\pi NN}^2 
\left[{e^{-m_0 r}\over r}\pm 2\,{e^{-m_c r}\over r}\right],
\end{equation}
where $m_c$ denotes the mass of the charged pion, $m_0$ that of the neutral
pion and the plus (minus) sign corresponds to isospin-singlet (-triplet) waves.
We use the same value of the $\pi N$ coupling as in the Nijmegen PWA's 
\cite{nijnn}, $f_{\pi NN}^2=0.075$. The waves are obtained by solving 
a Schr\"odinger equation with this potential. As expained in Ref.~\cite{friar}, 
relativistic effects can, to the order we are working, be absorbed into terms
in the TPE potential and a factor of $M/E$ multiplying the OPE potential.
With standing-wave boundary conditions appropriate to the $K$-matrix, the DW's
have the asymptotic form
\begin{equation}
\psi_{\scriptscriptstyle\rm OPE}(p,r)\,\rar_{r\rightarrow \infty}\,
{\sin(pr-L\pi/2)+\tan\delta_{\scriptscriptstyle\rm OPE}(p)
\cos(pr-L\pi/2)\over pr}.
\end{equation}
Near the origin, they behave like 
\begin{equation}\label{eq:asyr0}
\psi_{\scriptscriptstyle\rm OPE}(r)\propto {(pr)^L\over (2L+1)!!}
\qquad\mbox{as}\ r\rightarrow 0.
\end{equation}

If the residual scattering is weak, we can represent it using an effective 
theory based on a trivial fixed point \cite{bmr,bb}. The residual $K$-matrix 
$\tilde K(p)$ is then equal to the matrix element of an energy-dependent 
short-range potential, in the DW Born approximation. A simple $\delta$-function 
potential will have no effect since the waves with $L>0$ vanish at the origin. 
We could use an appropriate high derivative of a $\delta$-function to represent 
the short-range interactions but, for numerical implementation, it is more 
convenient to work with an energy-dependent $\delta$-shell 
potential, 
\begin{equation}\label{eq:vshort}
V_S(p,r)={[(2L+1)!!]^2\over 4\pi R_0^{2L+2}}\,\tilde V(p)\,\delta(r-R_0),
\end{equation}
Provided $R_0$ is chosen to be sufficient small that
the asymptotic form (\ref{eq:asyr0}) is valid, this form is numerically
equivalent to a derivative of a $\delta$-function. (In practice we take 
$R_0=0.1$~fm.) Here we have divided out a factor of $R_0^{-2L}$ so
that the strength $\tilde V(p)$ is independent of $R_0$ for small $R_0$. 
We have also divided out the numerical factor of $[(2L+1)!!]^2$ which is needed
to compensate for the smallness of the high partial waves at small radii, 
as can be seen in Eq.~(\ref{eq:asyr0}).

The residual scattering when leading-order OPE only is removed starts at 
order $Q^2$ in the chiral expansion. Equating the DW matrix
element of the short-range potential to $\tilde K(p)$, we find that its 
strength is given by
\begin{equation}
\tilde V^{(2)}(p)={R_0^{2L}\over [(2L+1)!!\,
\psi_{\scriptscriptstyle\rm OPE}(p,R_0)]^2}\,\tilde K(p).
\end{equation}
We can also remove the leading effects of order-$Q^{2,3}$ OPE and TPE by 
subtracting from $\tilde K(p)$ the matrix elements of these potentials.
The correction to OPE obtained by expanding the relativistic factor is
\begin{equation}
V^{(2)}_{\scriptscriptstyle\rm OPE}(r)=-{p^2\over 2M^2}\,
V^{(0)}_{\scriptscriptstyle\rm OPE}(r).
\end{equation}
The forms of the TPE potentials can be found in Refs.~\cite{kbw,nij99}.
The resulting residual scattering starts at order $Q^4$ and can be
described by a short-range potential with strength
\begin{equation}
\tilde V^{(4)}(p)={R_0^{2L}\over [(2L+1)!!\,
\psi_{\scriptscriptstyle\rm OPE}(p,R_0)]^2}\,\left(\tilde K(p)
-\langle\psi_{\scriptscriptstyle\rm OPE}(p)|V_{\scriptscriptstyle\rm OPE}^{(2)}
+V_{\scriptscriptstyle\rm TPE}^{(2,3)}|\psi_{\scriptscriptstyle\rm OPE}(p)
\rangle\right).
\end{equation}

The TPE potential at order $Q^3$ depends on the coefficients of three terms 
in order-$Q^2$ $\pi N$ Lagrangian. In the results presented here, we have used 
the values obtained by the Nijmegen group from their recent analysis of $NN$
data \cite{nij03}: $c_1=-0.76$~GeV$^{-1}$, $c_3=-4.78$~GeV$^{-1}$ and
$c_4=3.96$~GeV$^{-1}$. We have also checked that other sets of values,
for example those used in Refs.~\cite{kbw,egm1,nij99,egm2}, do not qualitatively
change our results.

The leading short-distance interaction in a partial wave with orbital angular
momentum $L$ can be represented by a contact interaction proportional to the 
$2L$-th derivative of a $\delta$-function. As mentioned above, we find it 
numerically more convenient to work with a $\delta$-shell interaction
proportional to $R_0^{-2(L+1)}\delta(r-R_0)$. In either case the leading
interaction is of order $Q^{2L}$ in the chiral expansion.
In momentum space it has a strength of order $\Lambda^{-2L+2}k^{2L}$ where 
$\Lambda$ is a scale associated with the underlying short-distance physics. 
One might expect that for a potential of ``natural'' strength, the scale 
$\Lambda$ should be of the order of the masses of the exchanged heavy mesons, 
at least 500~MeV. However, the rather strong $\pi N$ coupling introduces another,
significantly lower, scale of the order of 300~MeV \cite{BvK02}. The current
ChPT potentials do not include the $\Delta$, which also corresponds to a scale
of about 300~MeV.

The leading OPE potential is of order $Q^0$ and so it has the form 
$\Lambda^{-2}f_0(k/m_\pi)$ where $k$ is a generic momentum variable and
$f_0(x)$ is a dimensionless function of order 1. TPE contributions start at 
order $Q^2$ and have the form $\Lambda^{-4}m_\pi^2f_2(k/m_\pi)$. If, as in 
the present work, we include OPE and TPE interactions up to order $Q^3$, the 
omitted order-$Q^4$ forces arising from exchange of up to three pions are 
of the form $\Lambda^{-6}m_\pi^4f_4(k/m_\pi)$. 

The DW approach represents all higher-order effects in terms of short-range 
interactions. When we extract only the leading OPE potential, the effects of
order-$Q^2$ TPE contributions in the partial wave $L$ are replaced by an 
interaction with strength 
\begin{equation}\label{eq:v2size}
\tilde V^{(2)}\sim (L!)^2\Lambda^{-4}m_\pi^{(2-2L)}g_2(p/m_\pi), 
\end{equation}
where $p$ is the on-shell momentum and $g(x)$ is another dimensionless function 
of order 1. Here $2L$ powers of momenta have been extracted from $f_2(k/m_\pi)$ to 
form a projector onto the relevant partial wave. This projector involves $L$
derivatives of the initial and final wave functions, which leads to the 
numerical factor of $(L!)^2$. 

When we extract OPE and TPE interactions up to order $Q^3$, the residual 
interaction strength has the form
\begin{equation}\label{eq:v4size}
\tilde V^{(4)}\sim (L!)^2\Lambda^{-6}m_\pi^{4-2L}k^{2L}g_4(p/m_\pi). 
\end{equation}
In the case of a $D$-wave, this contains a momentum-independent term 
$\Lambda^{-6}$, which corresponds to a contact interaction with an unknown 
coefficient. 

By extracting the effects of the leading-order OPE and TPE forces, we expect
to remove the dominant energy dependence of the scattering amplitudes up
energies $T_{\rm lab}\sim 200$~MeV. Beyond that region, three-pion exchange
can start to contibute significantly, although calculations suggest that these 
forces are much smaller than other order-$Q^4$ contributions \cite{kais3}.

\section{Results}

\begin{figure}[t,b]
\includegraphics[width=18cm,keepaspectratio,clip]{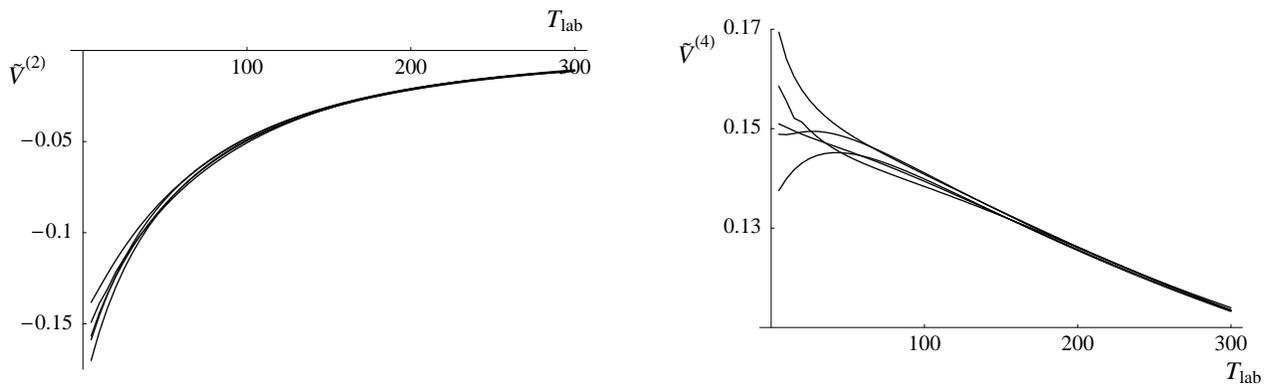}
%\vspace{0.5cm} \epsfxsize=18cm \epsffile{p2130.eps}
\caption{\label{fig:1D2} The short-range effective potential 
$\tilde V(p)$ in the $np$ $^1D_2$ partial wave, plotted in fm$^6$ against 
$T_{\rm lab}$ in MeV. The five curves correspond to the various Nijmegen PWA's 
and potentials. The left-hand panel shows the potential obtained when only 
leading-order OPE is removed; the right-hand one shows the result of also 
subtracting the order-$Q^{2,3}$ OPE and TPE potentials.}
\end{figure}

\begin{figure}[t,b]
\includegraphics[width=18cm,keepaspectratio,clip]{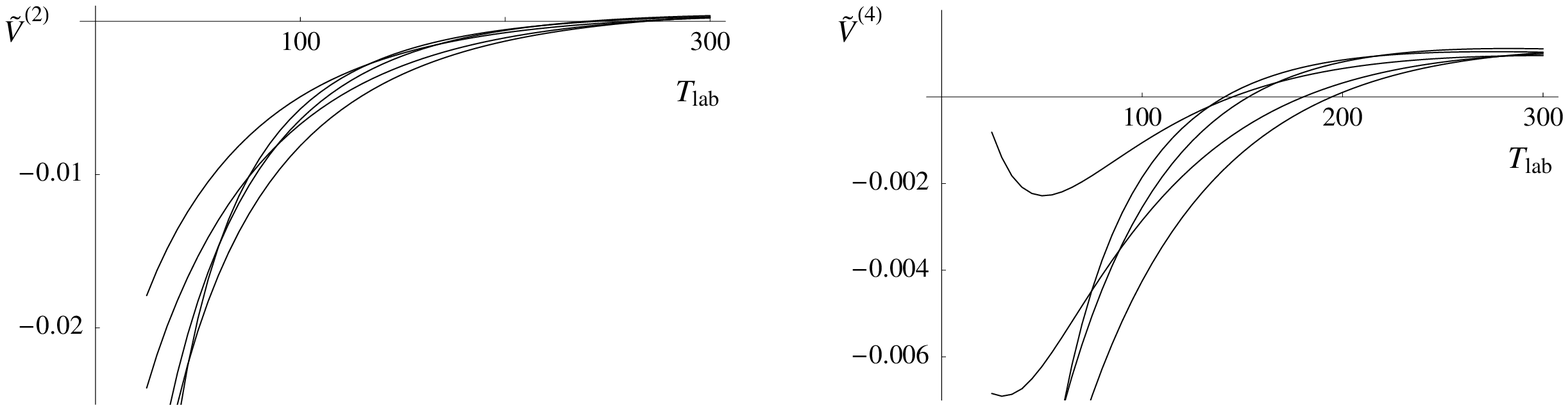}
%\vspace{0.5cm} \epsfxsize=18cm \epsffile{p3130.eps}
\caption{\label{fig:1F3} The short-range effective potential
in the $np$ $^1F_3$ partial wave, plotted in fm$^8$ against $T_{\rm lab}$ 
in MeV. For other details see the caption to Fig.~1.}
\end{figure}

\begin{figure}[t,b]
\includegraphics[width=18cm,keepaspectratio,clip]{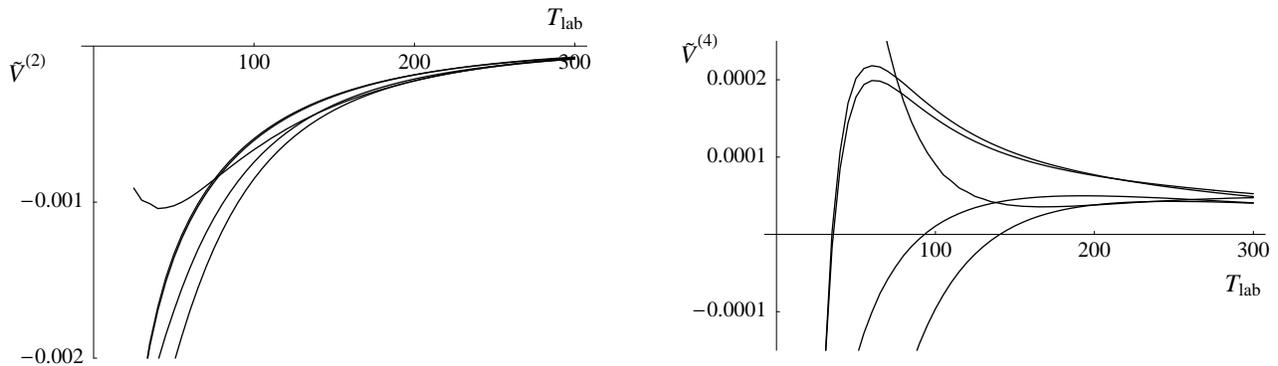}
%\vspace{0.5cm} \epsfxsize=18cm \epsffile{p4130.eps}
\caption{\label{fig:1G4}  The short-range effective potential
in the $np$ $^1G_4$ partial wave, plotted in fm$^{10}$ against $T_{\rm lab}$ 
in MeV. For other details see the caption to Fig.~1.}
\end{figure}

The main results of this analysis are shown in 
Figs.~\ref{fig:1D2}--\ref{fig:1G4}. (Note the differences in scale between
the two panels of each plot.) There are a couple of general lessons to 
be drawn from them before examining the individual waves in more detail.

First, it is clear that for energies below about 80 MeV there are
substantial differences between the various PWA's.
For the $^1F_3$ and $^1G_4$ waves in this region systematic artefacts of 
the different parametrisations completely dominate the differences 
between the empirical phase shifts and those from OPE 
plus TPE. Even in the range 150--250 MeV, the uncertainties 
are so large as to preclude much more than a rough estimate of the 
magnitude of the residual potential. Only in the $^1D_2$ wave does
one find consistent results over a wide range of energies, from about 50 MeV
upwards. One point to note about the $^1D_2$ and $^1G_4$ waves is that we
find no correlation among the deviations of the different PWA's from 
their common trend. This suggests that there is no systematic bias to the fits.

\begin{figure}[t,b]
\includegraphics[width=8cm,keepaspectratio,clip]{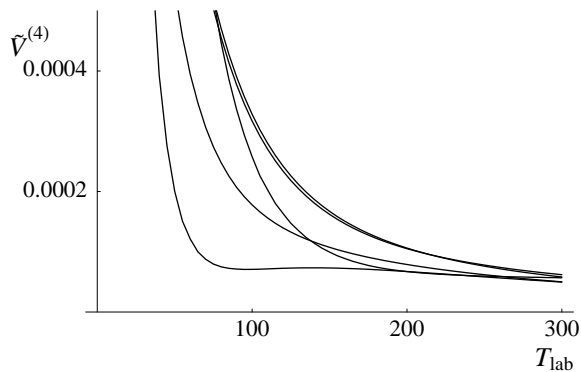}
%\vspace{0.5cm} \centerline{\epsfxsize=8cm \epsffile{d4030.eps}} 
\caption{\label{fig:1G4nr} The short-range effective potential
(in fm$^{10}$) in the $np$ $^1G_4$ partial wave, with leading-order OPE and 
order-$Q^{2,3}$ TPE removed but not the order $Q^2$ recoil correction to OPE. 
For other details see the caption to Fig.~1.}
\end{figure}

Second, it is important to use the ``correct" $\pi N$ coupling, namely the one 
assumed in the PWA, and to include the $M/E$ factor multiplying OPE (or at least 
the leading correction from it). If these are not done then the results show a 
strong systematic energy dependence at low energies. An example of this is
given in Fig.~\ref{fig:1G4nr}, where TPE has been subtracted but not the order-$Q^2$
correction to OPE. 
There are plots in Refs.~\cite{kbw,egm1,rich} comparing OPE plus TPE with PWA's
that all look similar to each other. However Ref.~\cite{kbw} includes the $M/E$ 
factor but uses a large value for the $\pi N$ coupling ($f_{\pi NN}^2=0.077$ as
opposed to the Nijmegen recommended value). In contrast 
Refs.~\cite{egm1,rich} omit the $M/E$ factor but use smaller $\pi N$ couplings. The
results of these two choices are quite similar for $^1F_3$ and $^1G_4$
scattering in the energy range 100--200 MeV, where difference from pure OPE are 
most visible. In both cases, the differences from the PWA's are larger than
those obtained when the consistent $\pi N$ coupling is used and recoil
corrections are included. Although these differences are similar in magnitude
to the systematic uncertainties in the current PWA's, they will become more 
significant when phase shifts from improved PWA's become available
\cite{nij99,nij03}.

In the $^1D_2$ wave, shown in Fig.~\ref{fig:1D2}, subtraction of the 
order-$Q^{2,3}$ terms may not have much effect on the size of the residual 
scattering, but it does dramatically reduce its energy dependence. There 
is a nearly 100\% change over the energy range 50 to 300 MeV if OPE alone 
is removed, but this is reduced to about 20\% when the order-$Q^{2,3}$ terms 
are subtracted. The typical size of the residual strength after removal of OPE,
$\tilde V^{(2)}$, is about 0.1~fm$^6$. Comparing this with Eq.~(\ref{eq:v2size})
we find that the corresponding scale $\Lambda$ is approximately 300~MeV, as expected
for pion-exchange forces. Ater subtracting OPE and TPE to order $Q^3$, the residual 
short-range potential is, to a very good approximation, linearly dependent on the 
energy. Its intercept occurs at about $\tilde V^{(4)}\simeq 0.16$~fm$^6$, 
which corresponds to a momentum scale of about 200~MeV in Eq.~(\ref{eq:v4size}). 
This is distinctly smaller than one would expect on grounds of ``naturalness". 
Its slope is ${\rm d}\tilde V^{(4)}/{\rm d}p^2\simeq 0.012$ fm$^8$, corresponding 
to a scale of about 370~MeV. 

This picture bears out what was found in Refs.~\cite{egm1,rich}, namely that 
one counterterm of 
order $Q^4$ is able to explain the bulk of the residual scattering after
all contributions up to order $Q^3$ have been removed.  One might worry
that those authors neglected the recoil correction to OPE. However in this 
channel the TPE contributions are so much larger that the neglect of recoil does 
not affect the results significantly. At least for the spin-singlet channels,
this analysis also removes the worries 
raised in Ref.~\cite{em2}: although there is one unnaturally large term
in the effective short-range potential, there is no evidence for a breakdown
of the chiral expansion, even up to energies of $\sim 300$ MeV.

In the $^1F_3$ and $^1G_4$ cases there are large differences between the various
PWA's for energies below about 150~MeV and so it is hard to draw such definite 
conclusions. Nonetheless it is clear that the residual scattering is much smaller 
after the order-$Q^{2,3}$ terms have been subtracted, as in the results of
Refs.~\cite{kbw,egm1}. The sizes of the residual scattering strengths, $V^{(2)}$
and $V^{(4)}$, at energies below 100~MeV in these waves correspond to scales in 
the region 300--400~MeV. 

Although the different PWA's show no overall bias in the $^1D_2$ and $^1G_4$ waves,
this is not the case in the $^1F_3$ wave. There, in contrast, all the residual
interactions show a significant downward curvature at low energies, even after
subtraction of the order-$Q^{2,3}$ terms. While this may just reflect the fact
this wave is poorly constrained by data (there are differences
between the Nijmegen \cite{nijnn} and VPI \cite{said} phase shifts for this wave) 
a more intriguing possibility is that it could be a signal of a long-ranged 
isospin-breaking effect. 

In this context, it should be noted that the 
isospin-triplet partial waves, such as $^1D_2$ and 
$^1G_4$, are fitted to $pp$ scattering data in the Nijmegen PWA's \cite{nijnn}. 
The corresponding $np$ results are then obtained by simply replacing the 
neutral-pion exchange in $pp$ by the relevant combination of neutral- and 
charged-pion exchange, as in Eq.~(\ref{eq:npope}). In contrast the 
isospin-singlet waves, such as $^1F_3$, have to be obtained from fitting 
$np$ data. A single value for the $\pi N$ coupling is used in all cases. 
The overall downward deviation in our results is consistent with
what would be expected if we had taken too high a $\pi N$ coupling
in the DW analysis of these waves. A similar pattern is also seen in the
$^1P_1$ wave. The size of the effect is compatible with what can be deduced 
from PWA's of $NN$, $\pi N$ or $N\overline N$ scattering \cite{srt}, 
however we would caution that all these effects are comparable to the 
uncertainties in the PWA's.  Other isospin-violating $NN$ interactions in the 
framework of ChPT have been discussed in Refs.~\cite{vkrfgs,fvkpc,egm3}.
The longest-ranged of these, and hence the most important at low-energies, are
the electromagnetic corrections to OPE \cite{vkrfgs}, but we find that the 
effect of subtracting the order-$\alpha$ corrections to OPE is very small for 
energies above 80 MeV. It will be interesting to see if this deviation remains 
when the DW method is applied to phase shifts from the newer chiral PWA's of 
the Nijmegen group \cite{nij99,nij03}.

\section{Conclusions}

We have presented here a method for extracting the effects of 
OPE and TPE from peripheral $NN$ phase shifts. A DW approach is used to remove 
the effects of simple OPE to all orders. TPE and recoil corrections to OPE 
are then subtracted pertubatively and we are then able to remove all 
contributions up to order $Q^3$. We have applied this technique to peripheral 
$NN$ scattering in spin-singlet waves, using phase shifts from various Nijmegen 
PWA's\cite{nijnn}. We find residual interactions which are consistent with 
the expectations for an effective field theory. In the $^1D_2$ wave, the energy 
dependence of the residual interaction is essentially linear up to nearly 300 MeV.
The systematic errors in the various PWA's are large in the $^1F_3$ and $^1G_4$ 
waves, making it hard to draw definite conclusions about the effectiveness of the 
theory in these cases. Nonetheless, the residual scattering in these waves is small 
after removal of all order-$Q^{2,3}$ terms. 

The momentum scales of the terms in the 
residual short-range potentials are, with one exception, 300~MeV or larger, as 
expected from the scale appearing the pion-exchange potentials. The exception is the 
leading term in the $^1D_2$ wave. This term is unnaturally large, corresponding to 
a scale of about 200 MeV. In the energy region up to $T_{\rm lab}\sim 200$~MeV,
where OPE and TPE forces are expected to dominate, there is no evidence of 
a breakdown of the chiral expansion, nor is it necessary to introduce any
additional regularisation in the spin-singlet channels.

In the isospin-singlet waves we find hints of isospin breaking 
in the $\pi N$ couplings, although these are very much at the limit what can be 
deduced given the systematic errors in the different PWA's. 

It will be very interesting to see the results of this DW method when it is 
applied to phase shifts from the newer Nijmegen PWA's 
\cite{nij99,nij03}. If these are suffiently well-determined at low-energies, it 
should be posssible to use the method to examine whether the results after
extraction of the order-$Q^4$ potential \cite{kais,em1,kais3,egm3} remain 
consistent with the chiral expansion.

\section*{Acknowledgments}
We are grateful to R. Timmermans for useful correspondance. MCB thanks 
K. Richardson and T. Barford for discussions about these ideas.
This work is supported by the EPSRC.

\end{document}